\documentclass[aps,twocolumn,showpacs,prl,preprintnumbers,amsmath,amssymb]{revtex4}
\newcommand{\lsim} 
 {\ \raise.35ex\hbox{$<$}\kern-0.75em\lower.5ex\hbox{$\sim$}\ }
\newcommand{\gsim}
 {\ \raise.35ex\hbox{$>$}\kern-0.75em\lower.5ex\hbox{$\sim$}\ }
\newcommand{\bras}[1]{\langle#1|}
\newcommand{\kets}[1]{|#1\rangle}

\newcommand{\means}[1]{\langle#1\rangle}

\def\journal #1#2#3#4{#1 {\bf #2}, #3 (#4)}
\def\PR{Phys.\ Rev.}

\def\PRB{Phys.\ Rev.\ B}
\def\PRE{Phys.\ Rev.\ E}
\def\PRL{Phys.\ Rev.\ Lett.}

\def\JPSJ{J.\ Phys.\ Soc.\ Jpn.}

\usepackage{graphicx}
\usepackage{dcolumn}
\usepackage{bm}
\usepackage{amsmath}
\usepackage{times}
\usepackage[usenames]{color}

\usepackage{ulem}

\begin{document}
\title{
Thermodynamics of Chiral Spin Liquids with Abelian and Non-Abelian Anyons}
\author{Joji Nasu$^{1}$ and Yukitoshi Motome$^{2}$} 
 \affiliation{$^{1}$Department of Physics, Tokyo Institute of Technology, Ookayama, 2-12-1, Meguro, Tokyo 152-8551, Japan,\\
$^{2}$Department of Applied Physics, University of Tokyo, Hongo, 7-3-1, Bunkyo, Tokyo 113-8656, Japan}
\date{\today}
\begin{abstract}
 Thermodynamic properties of chiral spin liquids are investigated for a variant of the Kitaev model defined on a decorated honeycomb lattice.
 Using the quantum Monte Carlo simulation, we find that the model exhibits a finite-temperature phase transition associated with the time reversal symmetry breaking, in both topologically trivial and nontrivial regions.
While changing the exchange constants, the phase transition changes from continuous to discontinuous one, apparently correlated with the change in the excitations from Abelian to non-Abelian anyons.
We show this coincidence by computing the topological quantities: the Chern number and the thermal Hall conductivity.
In addition, we find, as a diagnostic of the chiral spin liquids, successive crossovers with multi-stage entropy release above the critical temperature, which indicates that the hierarchical fractionalization of a quantum spin occurs differently between the two regions.
\end{abstract}

\pacs{75.10.Kt,75.10.Jm,75.30.Et}


\maketitle



%
%

%




Understanding of quantum spin liquids (QSLs) in magnets, where strong quantum fluctuations suppress magnetic ordering even at the lowest temperature ($T$), has been one of the most challenging subjects in strongly correlated electron systems~\cite{Anderson1973,Balents2010}. 
Among many possible realizations of QSLs, the chiral spin liquid (CSL), in which the time reversal symmetry is broken, has attracted considerable attention in not only condensed matter physics but also quantum information.
This is because it may have the excitations obeying the non-Abelian anyon statistics, which are utilized as the operators in topological quantum computing~\cite{Kitaev06}.
To explore this exotic state, quantum spin systems on geometrically frustrated lattices have been intensively studied thus far~\cite{Kalmeyer1987}.
For instance, the possibility of CSLs has been studied theoretically in the Heisenberg model on a kagome lattice~\cite{Yang1993,Gong2013,Bauer2014}.
Experimentally, a possible CSL was discussed for a metallic pyrochlore compound Pr$_2$Ir$_2$O$_7$~\cite{Machida2010}.

Besides the analyses of geometrically-frustrated quantum magnets, a class of the models that have the exact CSL ground states has opened a new avenue in the study of CSLs~\cite{Kitaev06,Yao2007,Schroeter2007}.
One of them was originally suggested by A. Kitaev~\cite{Kitaev06} and studied in detail by H. Yao and S. Kivelson~\cite{Yao2007}.
This model is a variant of the honeycomb Kitaev model, which is defined on a decorated honeycomb lattice, composed by extending each honeycomb lattice site to a triangle.
The exact solution of this model shows that the ground state accommodates two different types of the CSLs accompanied with Abelian and non-Abelian anyons as the elementary excitations.
Interestingly, the non-Abelian CSL has topologically nontrivial Majorana fermion bands with a chiral edge mode.

Although the exact solutions for the ground states serve as good references in the exploration of CSLs, it is a crucial issue how the CSLs behave against thermal fluctuations at finite $T$. 
As the discrete chiral symmetry is broken in the ground state, one expects a phase transition to the CSL at a finite $T$, even in two dimensions. 
An interesting issue is how the non-Abelian anyonic nature survives at finite $T$, since the robustness is crucial for the application in topological quantum computing~\cite{Chung2010}. 
It is also important to clarify the precursors of CSLs when approaching from high-$T$ paramagnet, as they may provide a clue for searching CSLs in experiments. 
However, finite-$T$ properties of CSLs remain elusive, mainly because of the lack of well-controlled theoretical methods for the quantum spin systems in question.
For instance, the conventional world-line quantum Monte Carlo (QMC) method, which is one of the most standard techniques for finite $T$, does not work efficiently due to the negative sign problem.

In this Letter, we elucidate the effect of thermal fluctuations on CSLs by unbiased numerical simulations.
We address this problem in a representative model that has the exact CSL ground state, a variant of the Kitaev model on a decorated honeycomb lattice mentioned above.
We compute the finite-$T$ properties of this model by adopting the QMC method formulated in the Majorana fermion representation, which does not suffer from the negative sign problem~\cite{Nasu2014,Nasu2015,Nasu2015b}.
We find that both topologically trivial and nontrivial CSLs exhibit a phase transition at a finite $T$ associated with the time reversal symmetry breaking. 
By changing the exchange parameters, the phase transition changes from second order to first order.
This appears to be related with the change of topological nature of the chiral phases, that is, from the Abelian to non-Abelian CSLs, which we corroborate by calculating the Chern number and the thermal Hall conductivity.
In addition, we clarify how the CSLs develop from the high-$T$ paramagnet while decreasing $T$;
there are several crossovers characterized by the multi-stage entropy release, reflecting hierarchical fractionalization of a quantum spin, which appears in a different way between the Abelian and non-Abelian regions.

\begin{figure}[t]
\begin{center}
\includegraphics[width=\columnwidth,clip]{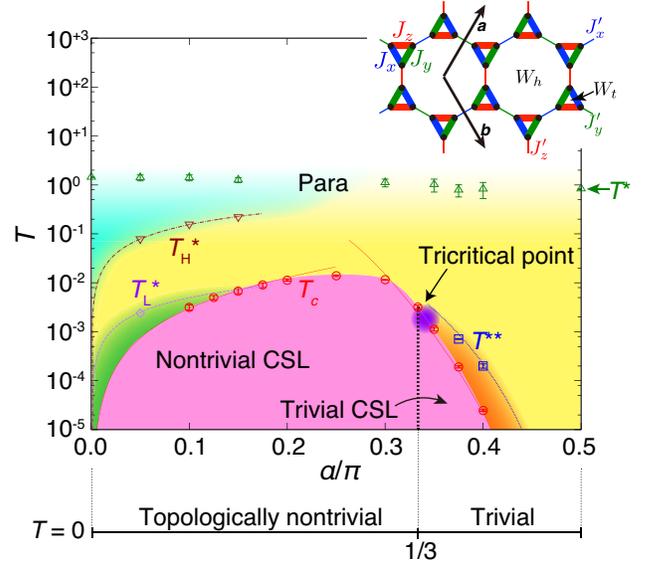}
\caption{(color online).
 Finite-$T$ phase diagram of the Kitaev model on a decorated honeycomb lattice.
 The ground-state phase diagram is also presented in the bottom of the figure. 
The lattice structure is depicted in the inset.
 The circles represent the phase transition temperature $T_c$.
 The deduced location of the tricritical point is also shown; for larger (smaller) $\alpha$, the transition is continuous (discontinuous).
The triangles, squares, diamonds, and inverted triangles represent the crossover temperatures, $T^*$, $T^{**}$, $T_{\rm L}^*$, and $T_{\rm H}^{*}$, respectively.
The solid and dotted lines in the large $\alpha$ region represent $T_c$ and $T^{**}$, respectively, determined by the MC simulation for the effective model for $J'/J\gg 1$.
The solid, dashed, and dashed-dotted lines in the small $\alpha$ region represent $T_c$, $T_{\rm L}^{*}$, and $T_{\rm H}^{*}$, respectively, obtained from the effective model for $J'/J\ll 1$.
 See the text for details.
 }
\label{phase}
\end{center}
\end{figure}

We consider a variant of the Kitaev model defined on a decorated honeycomb lattice, whose Hamiltonian is given by~\cite{Yao2007}
\begin{align}
 {\cal H}=-\sum_{\gamma =x,y,z}\sum_{\means{jk}_\gamma}J_\gamma \sigma_j^\gamma \sigma_k^\gamma
-\sum_{\gamma =x,y,z}\sum_{\means{jk}'_\gamma}J'_\gamma \sigma_j^\gamma \sigma_k^\gamma,\label{eq:1}
\end{align}
where $\sigma_{j}^\gamma$ represents the $\gamma$ component of the Pauli matrix describing an $S=1/2$ spin located at a site $j$.
The decorated honeycomb lattice consists of triangles connected by bonds;
the interaction $J_\gamma$ is defined for nearest-neighbor (NN) $\gamma$ bonds in the triangles, $\means{jk}_\gamma$, while $J'_\gamma$ for NN $\gamma'$ bonds connecting the triangles, $\means{jk}'_\gamma$, as shown in the inset of Fig.~\ref{phase}.
For simplicity, we assume $J=J_x=J_y=J_z$ and $J'=J_x'=J_y'=J_z'$ as in Ref.~\cite{Yao2007}, and introduce the parameter $\alpha$ so that $J=\cos\alpha$ and $J'=\sin\alpha$.

The model in Eq.~(\ref{eq:1}) is exactly solvable for the ground state~\cite{Yao2007}. 
There are two kinds of local conserved quantities: one is defined for each triangle $t$, $W_t=\prod_{j\in t}\sigma_j^{\gamma_j}$, and the other for each dodecagon $h$, $W_h=\prod_{j\in h}\sigma_j^{\gamma_j}$, where $\gamma_j$ denotes the bond not included in $t$ or $h$ among three NN bonds at the site $j$.
Note that $W_t$ changes its sign by the time reversal operation as it consists of the three products of Pauli matrices. 
The exact ground state is a CSL in which the time reversal symmetry is broken by a uniform alignment of $W_t$ as well as $W_h$. 
Remarkably, the ground state accommodates two topologically different CSLs depending on $\alpha$, as presented in the bottom of Fig.~\ref{phase}. 
The critical point is located at $\alpha=\alpha_c=\pi/3$: the ground state is a topologically trivial CSL for $\alpha > \alpha_c$, whereas it becomes topologically nontrivial for $\alpha < \alpha_c$.
The latter phase has the non-Abelian anyons in the excitation.

In order to compute thermodynamic properties of CSLs in this model, we adopt a QMC technique developed by the authors and their collaborator recently~\cite{Nasu2014,Nasu2015,Nasu2015b}.
The method is based on the Majorana fermion representation of the quantum spins via the Jordan-Wigner transformation~\cite{Yao2007,Chen2007,Feng2007,Chen2008}. 
In terms of the Majorana fermions, the model in Eq.~(\ref{eq:1}) is written as
\begin{align}
 {\cal H}&=iJ_x \sum_{\means{jk}_x}c_j c_k
-iJ_y\sum_{\means{jk}_y}c_j c_k
-iJ_z\sum_{\means{jk}_z}\eta_r c_j c_k\nonumber\\
&+iJ'_x \sum_{\means{jk}'_x}c_j c_k
-iJ'_y\sum_{\means{jk}'_y}c_j c_k
-iJ'_z\sum_{\means{jk}'_z}\eta_{r'} c_j c_k,\label{eq:2}
\end{align}
where $j<k$.
The operator $\eta_r=i \bar{c}_j \bar{c}_k$ defined on each $z$ and $z'$ bond is regarded as a $Z_2$ variable taking $\pm 1$, because it commutes with the Hamiltonian and $\eta_r^2=1$ ($r$ is the bond index).
Thus, the Hamiltonian describes the free Majorana fermions coupled with thermally-fluctuating $Z_2$ variables.
This representation enables the QMC simulation without the negative sign problem.
We carried out 40,000 MC steps for measurement after 10,000 MC steps for thermalization.
Moreover, we used the parallel tempering algorithm to avoid the slowing down at low $T$~\cite{Hukushima1996}: we prepared 16 replicas in each simulation.
We calculated the $6L^2$-site clusters up to $L=10$ with the twisted boundary condition~\cite{Nasu2014}.

Figure~\ref{phase} shows the phase diagram obtained by the QMC simulation.
We find that the model in Eq.~(\ref{eq:1}) exhibits a phase transition at a finite $T$, as expected for the discrete chiral symmetry breaking in the CSL phases. 
The critical temperature $T_c$ is determined by the specific heat $C_v$ and the chirality $\kappa$, as described below.
In addition to the transition, we find several crossovers as shown in the phase diagram; we will return to this point later.

\begin{figure}[t]
\begin{center}
\includegraphics[width=\columnwidth,clip]{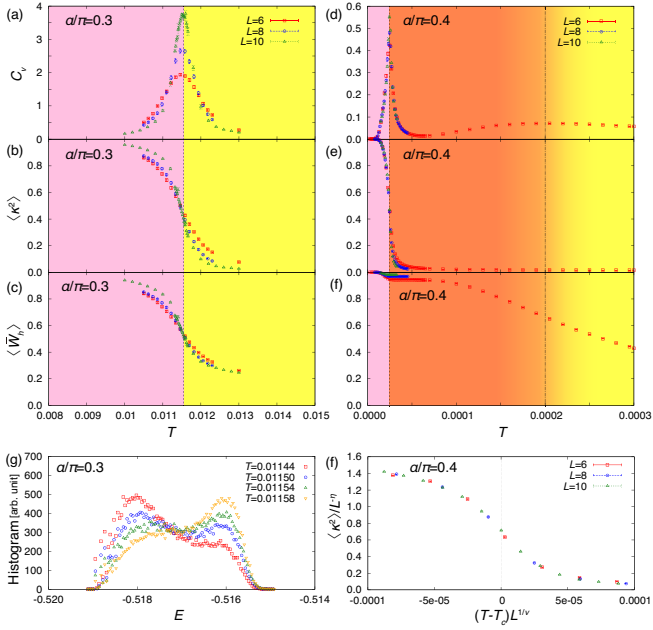}
\caption{(color online).
$T$ dependences of (a) the specific heat, (b) the mean square of the chirality $\kappa$, (c) the mean of $Z_2$ variables $W_h$ per dodecagon, $\bar{W}_h=\frac{1}{L^2}\sum_hW_h$, at $\alpha/\pi=0.3$.
The corresponding data at $\alpha=0.4$ are shown in (d), (e), and (f).
The vertical dashed (dashed-dotted) line indicates $T_c$ ($T^{**}$).
(g) Energy histograms at several $T$ in the vicinity of $T_c$ at $\alpha/\pi=0.3$.
(h) Scaling collapse for $\means{\kappa^2}$ with $1/\nu=1.09$ and $\eta=0.18$ at $\alpha/\pi=0.4$.
}
\label{Cv}
\end{center}
\end{figure}

QMC data for $C_v$ at $\alpha/\pi=0.3$ and 0.4 are shown in Figs.~\ref{Cv}(a) and~\ref{Cv}(d), respectively.
There is a sharp peak that grows with increasing the system size, indicating the phase transition.
We also show the data for the mean squares of the chirality defined as $\kappa=\frac{2}{L^2}\sum_{t} W_t$ in Figs.~\ref{Cv}(b) and~\ref{Cv}(e).
This quantity develops rapidly with decreasing $T$ at the peak temperature of $C_v$, which clearly indicates that the phase transition is associated with the time reversal symmetry breaking.

Interestingly, $T_c$ changes continuously while changing $\alpha$ as shown in Fig.~\ref{phase}, despite the topological change in the ground state at $\alpha=\alpha_c$. 
We, however, find that the nature of the phase transition changes in the vicinity of this point. 
To see this, we calculate the energy histogram at several $T$ near $T_c$. 
Figure~\ref{Cv}(g) shows the data at $\alpha/\pi=0.3$.
The histogram shows a double peak structure, which indicates that the phase transition is of first order. 
On the other hand, we cannot find such behavior at $\alpha/\pi=0.4$.
Instead, we show that the finite-size scaling collapse of $\langle\kappa^2\rangle$ works well for the $L=6$, 8, and 10 clusters as shown in Fig.~\ref{Cv}(h), suggesting that the phase transition is of second order at $\alpha/\pi =0.4$.
In the scaling collapse, the optimization is carried out by the Bayesian scaling analysis~\cite{Harada2011}
 and we obtain the critical exponents as $1/\nu=1.09(9)$ and $\eta=0.18(3)$.
 These exponents are close to those for the 2D Ising universality class, $\nu=1$ and $\eta=1/4$~\cite{comment}.
The results suggest the existence of the tricritical point between $\alpha/\pi=0.3$ and 0.4, which is close to $\alpha_c$, as shown in Fig.~\ref{phase}.

\begin{figure}[t]
\begin{center}
\includegraphics[width=\columnwidth,clip]{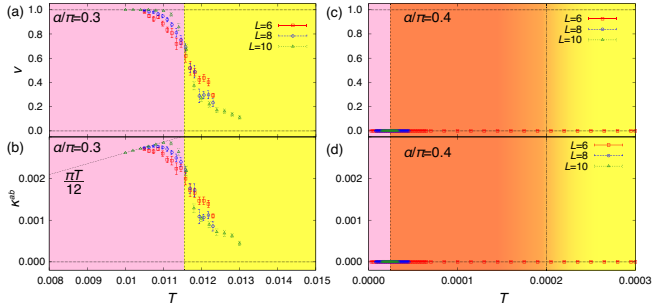}
\caption{(color online).
$T$ dependences of the absolute values of (a) the Chern number
$\nu$ and (b) the thermal Hall conductivity $\kappa^{ab}$ at $\alpha/\pi=0.3$.
 The vertical dashed line indicates $T_c$ and the dotted line in (b) represents $\pi T/12$.
Corresponding data at $\alpha/\pi=0.4$ are shown in (c) and (d).
}
\label{chiral}
\end{center}
\end{figure}

To examine the relation between the tricritical point and the topological nature of CSL phases below $T_c$, we calculate the topological quantities at finite $T$ by QMC. 
For this purpose, we first obtain the information of Majorana fermion bands.
In the ground state, the bilinear Majorana fermion Hamiltonian, which is given by Eq.~(\ref{eq:2}) with all $\eta_r=1$, is diagonalized as 
$
{\cal H}=\sum_{\bm{k}}\bm{c}_{\bm{k}}^\dagger H_{\bm{k}}\bm{c}_{\bm{k}}=\sum'_n \sum_{\bm{k}} \varepsilon_{n\bm{k}} (2f_{n\bm{k}}^\dagger f_{n\bm{k}}-1)
$,
where $\bm{c}_{\bm{k}}$ is a set of the Fourier transforms of $c_j$ in a unit cell and $H_{\bm{k}}$ is the Bloch Hamiltonian.
Here, we introduce $f_{n\bm{k}}$ and $f_{n\bm{k}}^\dagger$ as fermion operators for the $n$-th band. 
The summation $\sum'_n$ is taken only for the states with positive eigenvalues $\varepsilon_{n\bm{k}}$. 
In the finite-$T$ calculations, we extend these definitions straightforwardly by considering the $L_k^2$ supercell of $6L^2$ cluster obtained in the QMC simulation. 

First, we compute the Chern number, which is nonzero in the topologically nontrivial CSL ground state for $\alpha < \alpha_c$~\cite{Yao2007,Shi2010}. 
We extend the definition to finite $T$ as
$
\nu(T,\{\eta_r\}) = \frac{4\pi}{V} \sum_{n,\bm{k}}f_{\rm F}(E_{n\bm{k}})\sum_{m\neq n} 
{\rm Im} \frac{\bras{u_{n\bm{k}}}v_a\kets{u_{m\bm{k}}}
\bras{u_{m\bm{k}}}v_b\kets{u_{n\bm{k}}} }{(\varepsilon_{n\bm{k}}-\varepsilon_{m\bm{k}})^2+\delta^2}
$, 
where $V=L_k^2$, $f_{\rm F}$ is the Fermi distribution function, and the one-particle energy is given by $E_{n\bm{k}}=2|\varepsilon_{n\bm{k}}|$; 
$\kets{u_{n\bm{k}}}$ is the eigenvector for $H_{\bm{k}}$ for a given configuration of $\{\eta_r\}$; 
$v_l=\partial H_{\bm{k}}/\partial k_l$ ($l=a,b$), where $k_a$ and $k_b$ are the projections of $\bm{k}$ onto the reciprocal primitive vectors (see the inset of Fig.~\ref{phase}), and we assume these vectors are orthogonal~\cite{Shi2010}.
In the present calculation, we take $L_k=10$ and $\delta=10^{-2}$; we confirmed the convergence with respect to $L_k$ and $\delta$. 
Figure~\ref{chiral}(a) shows the QMC data for $\nu(T)=\means{|\nu(T,\{\eta_r\})|}$, which is computed for 100 samples during 40,000 MC steps.
At $\alpha/\pi=0.3$, $\nu(T)$ decreases rapidly near $T_c$ while increasing $T$, and the change becomes sharper for larger $L$, reflecting the first-order nature of the transition. 
This suggests a discontinuous change of $\nu(T)$ from $1$ to $0$ at $T_c$ in the thermodynamic limit.
In contrast, $\nu(T)$ is always zero at $\alpha/\pi=0.4$, as shown in Fig.~\ref{chiral}(c).

Next, we compute the thermal Hall conductivity, which reflects the topological nature of the excitations, since the heat is carried by the itinerant Majorana fermions $\{c_j\}$. 
The thermal Hall conductivity $\kappa^{ab}(T)$ is evaluated in a similar way to the Chern number:
$\kappa^{ab}(T)=\means{|\kappa^{ab}(T,\{\eta_r\})|}$, where
$\kappa^{ab}(T,\{\eta_r\})=\frac{T}{V}\sum_{n,\bm{k}}c_2(E_{n\bm{k}})\sum_{m\neq n} {\rm Im}\frac{\bras{u_{n\bm{k}}}v_a\kets{u_{m\bm{k}}}
\bras{u_{m\bm{k}}}v_b\kets{u_{n\bm{k}}} }{(\varepsilon_{n\bm{k}}-\varepsilon_{m\bm{k}})^2+\delta^2}$ with
$c_2(E_{n\bm{k}})=\int_{E_{n\bm{k}}}^\infty dE (\beta E)^2 \{-f_{\rm F}'(E)\}$~\cite{Luttinger1964,Kane1996,Cappelli2001,Qin2011,Sumiyoshi2013}. 
Figure~\ref{chiral}(b) shows the QMC results. 
Similarly to the Chern number $\nu(T)$, $\kappa^{ab}(T)$ sharply changes near $T_c$ at $\alpha/\pi=0.3$. 
At low $T$, $\kappa^{ab}$ shows $T$-linear behavior with the quantized coefficient as $\kappa^{ab}=\pi T/12$.
On the other hand, $\kappa^{ab}$ is always zero at $\alpha/\pi=0.4$, as shown in Fig.~\ref{chiral}(d).

Thus, both topological quantities $\nu$ and $\kappa^{ab}$ behave differently at $\alpha/\pi=0.3$ and $0.4$: they become nonzero below $T_c$ for the former, while always zero for the latter. 
The results suggest that the topological nature of CSL changes between the both sides of the tricritical point at $\sim \alpha_c$ separating the continuous and discontinuous phase transitions.

In addition to the change of the nature of phase transition and ordered phase, we find that the paramagnetic phase above $T_c$ also shows distinct behavior for $\alpha < \alpha_c$ and $\alpha > \alpha_c$: several crossovers appear in a different way, as shown in Fig.~\ref{phase}.
First of all, the system exhibits a crossover at $T=T^*\sim 1$ in both regions, which is almost constant while changing $\alpha$. 
In addition, for $\alpha < \alpha_c$, we obtain two crossovers at $T=T_{\rm H}^*$ and $T_{\rm L}^*$ ($T_{\rm H}^* > T_{\rm L}^*$). 
On the other hand, for $\alpha > \alpha_c$, the system exhibits a crossover at $T=T^{**}$ below $T^*$.
These crossovers are signatures of thermal fractionalization of a quantum spin into Majorana fermions, as explained below.

\begin{figure}[t]
\begin{center}
\includegraphics[width=0.9\columnwidth,clip]{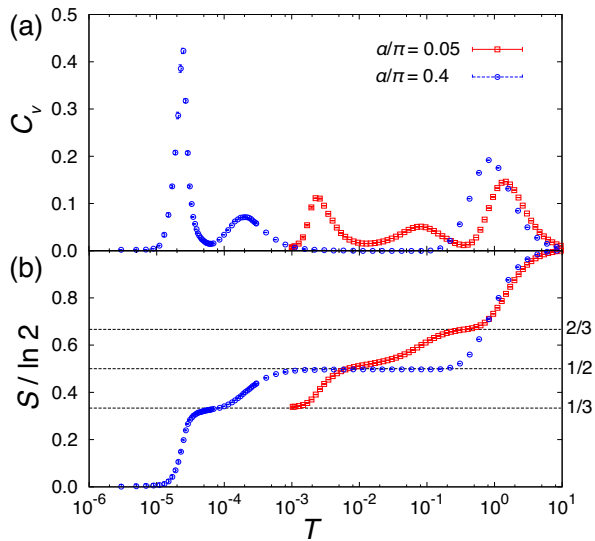}
\caption{(color online).
 $T$ dependences of (a) the specific heat and (b) the entropy per site at $\alpha/\pi=0.05$ and 0.4 in the $L=6$ cluster.
}
\label{CvS}
\end{center}
\end{figure}

First, we discuss the crossovers for $\alpha > \alpha_c$. 
Figure~\ref{CvS}(a) displays $T$ dependence of $C_v$ at $\alpha/\pi =0.4$. 
The data show three peaks: a broad peak appears at $T=T^{**}\sim 2\times 10^{-4}$, in between $T_c \sim 2\times 10^{-5}$ and $T^* \sim 1$. 
The high-$T$ crossover at $T^*$ comes from the kinetic motion of itinerant Majorana fermions $\{c_j\}$ in Eq.~(\ref{eq:2}) on the inter-triangle $J'$ bonds, which corresponds to the development of spin-spin correlations on the dimers.
The entropy of $\frac12 \ln 2$ is released in this crossover, as shown in Fig.~\ref{CvS}(b). 
On the other hand, the low-$T$ crossover at $T^{**}$ and the phase transition at $T_c$ are caused by localized Majorana fermions: $T^{**}$ corresponds to the coherent alignment of local conserved quantities $\{ W_h \}$ for dodecagons as shown in Fig.~\ref{Cv}(f), while $T_c$ corresponds to the other local conserved quantities $\{ W_t \}$ for triangles [or equivalently the chiral order parameter shown in Fig.~\ref{Cv}(e)]. 
Correspondingly, the entropy is released by $\frac16 \ln2$ and $\frac13 \ln2$, as shown in Fig.~\ref{CvS}(b).
This multi-state entropy release is explained by considering the limit of $\alpha/\pi \to 0.5$ ($J'\gg J$), where an effective Hamiltonian is obtained by the perturbation in terms of $J/J'$~\cite{Dusuel2008}. 
Investigating the effective model by a classical Monte Carlo simulation, we obtained the asymptotic behavior of $T^{**}$ and $T_c$, both of which well agree with the QMC results for the model in Eq.~(\ref{eq:2}), as shown in Fig.~\ref{phase}~\cite{Nasu_unpublished}.
Interestingly, $T^{**}$ appears to merge into $T_c$ in the vicinity of the tricritical point; for a smaller $\alpha$, 
the coherent alignments of $\{ W_h \}$ and $\{ W_t \}$ take place simultaneously in the first-order transition at $T_c$, as shown in Figs.~\ref{Cv}(b) and~\ref{Cv}(c).

Next, let us discuss the crossovers for $\alpha < \alpha_c$. 
As shown in the data at $\alpha/\pi = 0.05$ in Fig.~\ref{CvS}, the system shows four-stage entropy release in this region: 
$\frac13 \ln2$ at $T=T^*\sim J\sim 1$, $\frac16 \ln2$ at $T_{\rm H}^*\sim 10^{-1}$, $\frac16 \ln2$ at $T_{\rm L}^* \sim 2\times 10^{-3}$, and $\frac13 \ln2$ at $T_c < 10^{-3}$ ($T_c$ is unreachable because of the severe slowing down in QMC). 
Similar to the case with $J'\gg J$, the crossover at $T^*$ comes from itinerant Majorana fermions, corresponding to the spin-spin correlations are developed on the intra-triangle bonds.
On the other hand, the two crossovers at $T_{\rm L}^*$ and $T_{\rm H}^*$ and the phase transition at $T_c$ are explained by an effective model in the limit of $\alpha \to 0$ ($J\gg J'$), as follows.
The effective Hamiltonian is described by the pseudo spin $\bm{\tau}$ and $W_t$, which represent the four-fold degeneracy on each triangle at $J'=0$~\cite{Dusuel2008}.
It has a similar form to the Kitaev model on a honeycomb lattice in the magnetic field $h_{\rm eff}$, whose ground state is topologically nontrivial and excitations are non-Abelian anyon~\cite{Kitaev06}.
This consideration, together with the numerical results for the Kitaev model~\cite{Nasu2015b}, suggests that the system shows two crossovers at $T_{\rm L}^*\simeq 0.048J_{\rm eff}$ and $T_{\rm H}^*\simeq 0.15J_{\rm eff}$, where $J_{\rm eff}=J'/3$, as a consequence of thermal fractionalization of pseudo spin $\bm{\tau}$ into two kinds of Majorana fermions.
Moreover, $T_c$ is also suggested as $T_c\propto J'^2/J$ from the effective model.
The asymptotic behaviors are shown in Fig.~\ref{phase} by the dashed, dashed-dotted, and solid lines, which well agree with the QMC data for the model in Eq.~(\ref{eq:2}).
Interestingly, this multi-stage entropy release indicates that the fractionalization of a quantum spin occurs differently in this non-Abelian region: 
the original spin $\bm{\sigma}$ is first fractionalized into a pseudospin $\bm{\tau}$ and $W_t$, and furthermore, $\bm{\tau}$ is fractionalized into two Majorana fermions.

In summary, we have clarified the finite-$T$ properties of both topologically nontrivial and trivial CSLs realized in the Kitaev model on the decorated honeycomb lattice, by performing the QMC simulation in the Majorana fermion representation.
We revealed that both CSLs exhibits the phase transition with chiral symmetry breaking at finite $T$, whose critical temperatures $T_c$ seemingly meet with each other at the tricritical point.
We corroborate this by computing the topological quantities such as the Chern number and the thermal Hall conductivity.
We also find that the system exhibits several crossovers above $T_c$, reflecting the hierarchical fractionalization of a quantum spin into Majorana fermions.
The present results promote understanding of both the CSL phases with Abelian and non-Abelian excitations at finite $T$, and furthermore, the diagnostic of them in the high-$T$ paramagnetic phase, which will be useful for the experimental exploration of CSLs in quantum magnets.

\begin{acknowledgments}
We thank M. Udagawa for fruitful discussion. This work is supported by Grant-in-Aid for Scientific Research, the Strategic Programs for Innovative Research (SPIRE), MEXT, and the Computational Materials Science Initiative (CMSI), Japan.
Parts of the numerical calculations are performed in the supercomputing systems in ISSP, the University of Tokyo.
\end{acknowledgments}


\begin{thebibliography}{99} 

\bibitem{Anderson1973}
P. W. Anderson,
\journal{Mater. Res. Bull.}{8}{153}{1973}.

\bibitem{Balents2010}
L.~Balents,
\journal{Nature}{464}{199}{2010}.

\bibitem{Kitaev06}
A.~Kitaev,
\journal{Ann. Phys.}{321}{2}{2006}.


\bibitem{Kalmeyer1987}
V. Kalmeyer and R. B. Laughlin,
\journal{\PRL}{59}{2095}{1987}.

\bibitem{Yang1993}
K. Yang, L. K. Warman, and S. M. Girvin,
\journal{\PRL}{70}{2641}{1993}.

\bibitem{Gong2013}
S. Gong, W. Zhu, and D. N. Sheng,
\journal{Sci. Rep.}{4}{6317}{2014}.

\bibitem{Bauer2014}
B. Bauer, L. Cincio, B. P. Keller, M. Dolfi, G. Vidal, S. Trebst, and A W. W. Ludwig, 
\journal{Nat. Commun.}{5}{5137}{2014}.

\bibitem{Machida2010}
Y. Machida, S. Nakatsuji, S. Onoda, T. Tayama, and T. Sakakibara, \journal{Nature}{463}{210}{2010}.



\bibitem{Yao2007}
H. Yao and S. A. Kivelson,
\journal{\PRL}{99}{247203}{2007}.


\bibitem{Schroeter2007}
D. F. Schroeter, E. Kapit, R. Thomale, and M. Greiter,
\journal{\PRL}{99}{097202}{2007}.


\bibitem{Chung2010}
S. B. Chung, H. Yao, T. L. Hughes, and E.-A. Kim,
\journal{\PRB}{81}{060403}{2010}.



\bibitem{Nasu2014}
J. Nasu, M. Udagawa, and Y. Motome, 
\journal{\PRL}{113}{197205}{2014}.

\bibitem{Nasu2015}
J. Nasu, M. Udagawa, and Y. Motome, 
\journal{J. Phys.: Conf. Ser.}{592}{012115}{2015}.

\bibitem{Nasu2015b}
J. Nasu, M. Udagawa, and Y. Motome, 
arXiv:1504.1259.



\bibitem{Chen2007}
H.-D. Chen and J. Hu,
\journal{\PRB}{76}{193101}{2007}.

\bibitem{Feng2007}
X.-Y. Feng, G.-M. Zhang, and T. Xiang, 
\journal{\PRL}{98}{087204}{2007}.

\bibitem{Chen2008}
H.-D. Chen, and Z. Nussinov, 
\journal{J. Phys. A Math. Theor.}{41}{075001}{2008}.



\bibitem{Hukushima1996}
K. Hukushima and K. Nemoto,
\journal{\JPSJ}{65}{1604}{1996}.


\bibitem{Harada2011}
K. Harada,
\journal{\PRE}{84}{056704}{2011}.


\bibitem{comment}
The slight deviation of $\eta$ may be due to the limited system sizes used in the scaling. 
Indeed, we obtained better agreement for larger system sizes for the effective model in the isolated dimer limit, $J'\gg J$. The details will be published elsewhere.



\bibitem{Shi2010}
X.-F. Shi, Y. Chen, and J. Q. You,
\journal{\PRB}{82}{174412}{2010}.


\bibitem{Luttinger1964}
J. M. Luttinger,
\journal{\PR}{135}{A1505}{1964}.

\bibitem{Kane1996}
C. L. Kane and M. P. A. Fisher,
\journal{\PRB}{55}{15832}{1997}.

\bibitem{Cappelli2001}
A. Cappelli, M. Huerta, and G. R. Zemba,
\journal{Nucl. Phys. B}{636}{568}{2001}.

\bibitem{Qin2011}
T. Qin, Q. Niu, and J. Shi,
\journal{\PRL}{107}{236601}{2011}.

\bibitem{Sumiyoshi2013}
H. Sumiyoshi and S. Fujimoto,
\journal{\JPSJ}{82}{023602}{2013}.



\bibitem{Dusuel2008}
S. Dusuel, K. P. Schmidt, J. Vidal, and R. L. Zaffino,
\journal{\PRB}{78}{125102}{2008}.

        
\bibitem{Nasu_unpublished}
J. Nasu and Y. Motome, unpublished.

\end{thebibliography}


\end{document}